# Limited-angle TOF-PET for intraoperative surgical applications: proof of concept and first experimental data

S Sajedi[1], L Bläckberg[1], S Majewski[2], H Sabet[1]
[1]Department of Radiology, Massachusetts General Hospital, Harvard Medical School
[2]Radiology Department, University of California at Davis
E-mail: HSabet@MGH.Harvard.edu

**Abstract:** Intraoperative Gamma Probe (IPG) remains the current gold standard modality for sentinel lymph node identification and tumor removal in cancer patients. However, even alongside the optical dyes they do not meet with <5% false negative rates (FNR) requirement, a key metric suggested by the American Society of Clinical Oncology (ASCO). We are aiming to reduce FNR by using time of flight (TOF) PET detector technology in the limited angle geometry system by using only two detector buckets in coincidence, where one small-area detector is placed above the patient and the other with larger detection-area, placed just under the patient bed. For proof of concept, we used two Hamamatsu TOF PET detector modules (C13500-4075YC-12) featuring 12×12 array of 4.14×4.14×20 mm$^3$ LFS crystal pixels with 4.2 mm pitch, one-one coupled to silicon photomultiplier (SiPM) pixels. Detector coincidence timing resolution (CTR) measured 271 ps FWHM for the whole detector. We 3D printed lesion phantom containing spheres with 2-10 mm in diameter, representing lymph nodes, and placed it inside a 10-liter warm background water phantom. Experimental results show that with sub-minute data acquisition, 6 mm diameter spheres can be identified in the image when a lesion phantom with 10:1 activity ratio to background is used. Simulation results are in good agreement with the experimental data, by resolving 6 mm diameters lesions with 60 seconds acquisition time, in 25 cm deep background water phantom with 10:1 activity ratio. As expected, the image quality improves as the CTR improves in the simulation, and with decreasing background water phantom depth or lesion to background activity ratio, in the experiment. With the results presented here we conclude that limited angle TOF PET detector is a major step forward for intraoperative applications in that, improved lesion detectability is beyond what the conventional Gamma- and NIR-based probes could achieve.

## 1 Introduction

Intravenous (IV) injection of $^{18}$F-fludeoxyglucose (FDG) and Positron Emission Tomography (PET) imaging is considered as a standard in tumor detection and disease staging for cancer patients. However, it has been shown that this approach is only reliable for detecting tumors larger than 1 cm using the conventional whole-body PET (WB-PET) scanner, and thus will often miss small-size tumors and cancerous lymph nodes [1]. The key reasons for this poor diagnostic power are the relatively low spatial resolution (4-5 mm FWHM), low standardized uptake value (SUV) associated with $^{18}$F -FDG, as well as the inherently low sensitivity (~1%) of a conventional WB-PET scanner. These limitations have led to the development of intraoperative devices where detectors are placed close to the tissue of interest to increase the solid angle and hence the detection efficiency.

The sentinel lymph node (SLN) concept is based on the orderly progression of tumor cells within the lymphatic system. The SLN will be the first to contain metastasis, and biopsy will accurately predict regional nodal status [2]. The current standard of care in a breast cancer patient is to use both injection with blue dye (followed by direct vision inspection during surgery) and with peri-tumoral injection of radiopharmaceuticals for intraoperative identification of SLNs, using a gamma probe [3-8]. It has been shown that histological evaluation of one or more SLNs increases the accuracy of histopathologic staging of axilla in patients with breast cancer. Histological examination of LNs is important in determining the metastatic involvement; however, this examination requires node dissection and is associated with immediate and late postsurgical complications, especially lymphedema [9]. It should be noted that a thorough histological evaluation of 5 or more nodes is impractical or even impossible in the current standard of care. Also, the number of sections per node undergoing histological examination is limited (typically <4) which can lead to under-sampling of the involved nodes [10, 11] while having its associated complications. Furthermore, this approach has a significant false negative rate associated with cancer cells that pass downstream in the lymphatic system and miss the sentinel nodes [12]. Guidelines from the ASCO [13] states that surgical practices should aim for 85% identification rate and less than 5% false negative rates (FNR), which is not met by the current standard of care [14]. This indicates that there is an acute clinical need for tools that allow to efficiently and accurately examine a large enough number of nodes intra-operatively, before resection.

The most commonly used technology to identify SNL is gamma probe that, as mentioned above, have relatively high false negative rate. Detection sensitivity of these devices are also dependent on the tumor location and its performance is poor near the injection site and in deeply positioned tumors. It is noteworthy that the SNL identification is currently used as standard of care in breast and skin cancer patients. Another factor that hampers the usefulness of the gamma probes is that they cannot provide depth information unless they are rotated with respect to the FOV in multiple angular positions to estimate tissue depth.

The other alternative in radiation-based intraoperative identification of SNL is the use of short-range positron particles of PET tracers such as FDG. There is a large body of work dedicated to development of positron probes for intraoperative use [15-20]. Note that IV-injection of patients with FDG has its own pros and cons compared with intra-tumoral injection of $^{99m}$Tc colloids.

The advent of time of flight (TOF) PET has shown great promise in increasing image signal to noise and patient dose reduction [21, 22]. TOF-PET with high coincidence timing resolution (CTR) has enabled high quality PET imaging where TOF PET recently found its way in to main manufacturers' product line. Thanks to the most recent developments in high CTR PET detectors (down to ~200 ps FWHM), diagnostic PET imaging with limited angle detector coverage is now an active area of research [21]. Encouraged by these developments, our group is now involved in implementing TOF-PET for intraoperative imaging as a strong alternative to intraoperative gamma probes, positron probes, 3D gamma cameras [23], and preoperative PET [24, 25]. In this work, we lay down our recent proof of concept work for intraoperative TOF-PET with simulations and pilot experimental data.

## 2 Materials and methods

### 2.1 Simulation setup

In our recent work we have shown through GATE (GEANT4) simulation, that by bringing detector modules close to the patient, detector solid angle and thus the geometrical sensitivity rapidly increase even with small number of detector modules [15, 22]. Our previous studies showed that by using ~1/3$^{rd}$ of detector modules from a whole-body PET and rearranging them in a limited-angle body-contouring configuration (see Fig. 1), one can preserve system sensitivity and resolution [15, 22]. With these encouraging initial results, we further included flat panel detector geometries where the detector modules are placed in two parallel planes one right above the patient body and one below the surgery bed.

In this work we simulated a practical first configuration that consists of a commercially available TOF-PET detector technology from Hamamatsu Photonics used in a dual flat panel detector configuration. The simulation setup models C13500-4075YC-12 TOF-PET detectors (Hamamatsu Photonics, Japan). Each detector module is comprised of 12×12 array of 4.14×4.14×20 mm$^3$ lutetium fine silicate (LFS) crystal pixels with 4.2 mm pitch one-one coupled with silicone photomultiplier (SiPM) pixels. The overall active area of the detector module is 51×51 mm$^2$. External dimension of the module including light-tight housing is 53×53 mm$^2$.

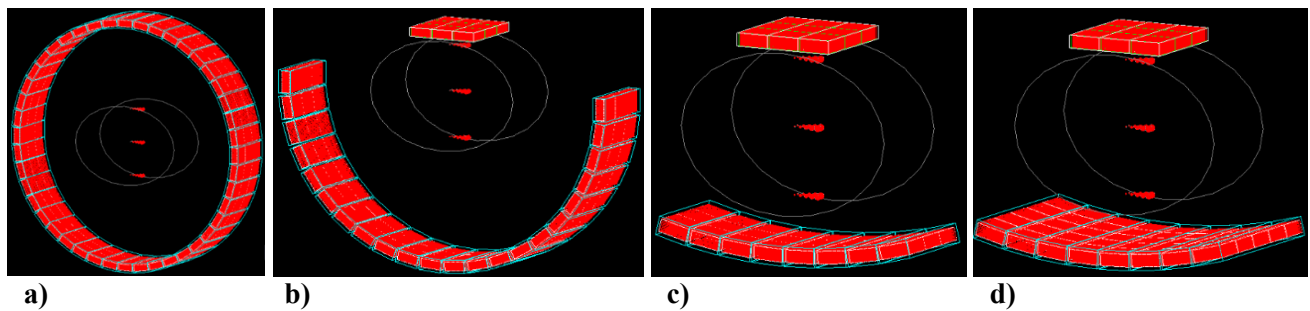

a)           b)           c)           d)

Fig. 1. GATE render of various intraoperative PET system geometries showing evolution of WB-PET to intraoperative PET. a) A whole-body PET with 76-cm diameter and 15 cm axial length. There are 44, 5.1×5.1 cm$^2$ detector modules per ring and 3 rings in the gantry totaling 132 modules. b) Only bottom half of the WB-PET in coincidence with 15×15 cm$^2$ detector panel. Total module numbers are 75 (9+132/2) c) first schematic of possible intraoperative PET with 8×3 modules under patient and 3x3 modules above patient (total 33). d) Intraoperative PET in (c) evolves to geometry with increased detector coverage while maintaining small footprint above the patient where surgical procedure takes place. In (d) there are 40 (8×5) and 9 (3×3) detector modules under and above the patient, respectively (total of 49) [22].

### 2.1.1 Phantom

A rectangular container measuring 24.5×13.5×25 (L×W×H) cm$^3$ was simulated as a slice of patient body within the scanner FOV. The container was filled with 5.3 kBq/cc with back-to-back gamma source. The background tracer concentration is based on a 10 mCi injection of $^{18}$F-

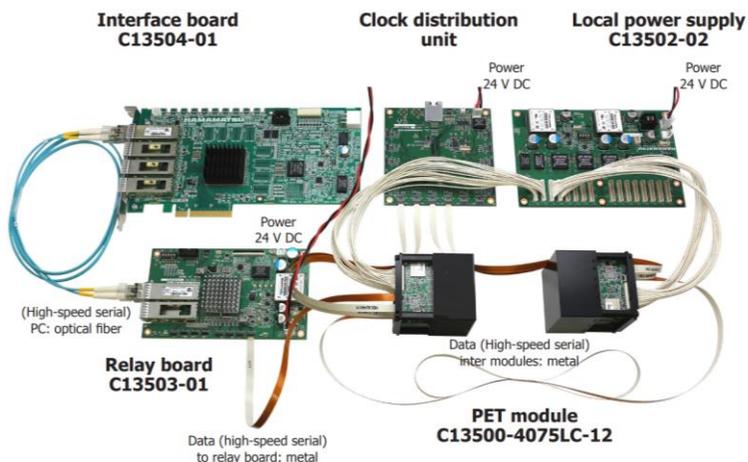

Fig. 2. Hamamatsu TOF-PET two-module evaluation kit with readout electronics and cabling.

FDG in to a 70 kg human. Hot spheres with 2, 4, 6, 8, and 9.5 mm diameter in two rows with 10:1 tracer uptake compared to background were used as tumor and lymph nodes. The $^{18}$F activities of the hot spheres were calculated at 222, 1776, 5994, 14208, and 23790 bq. Center to center distances of the hot spheres were kept at 1 cm. Hot spheres were positioned in the rectangular phantom with their centers positioned 2 cm below the phantom surface. GATE render of the hot lesion phantom is shown in Fig. 3 (top-right).

### 2.1.2 Detector placement and data acquisition

We simulated two detector panels with different numbers of detector modules per panel. In the first study the detector panel placed atop the patient, contains of only two modules while the one underneath the patient bed contains a 3×3 array of same size modules. The number of detectors per panel is selected based on the acquisition time limitations that we faced with only having two Hamamatsu TOF-PET detectors and the use of short half-life $^{18}$F isotope. The face to face distance between the two detector panels was 27 cm. Simulation was performed across 20 CPU threads with 9 seconds per thread using random engine with automatic seed. Detectors were modeled with 191 ps single module time blurring resulting in 271 ps CTR and %16.7 energy blurring to be consistent with the measured values (see sections 2.2.6 and 2.2.7) for the Hamamatsu TOF-PET detectors.

## 2.2 Experimental setup

### 2.2.1 Demo TOF-PET unit

Hamamatsu provided us with TOF-PET demo unit that included two C13500-4075YC-12 detector modules, a clock distribution board, a power supply board, a relay board, a PCIe interface board, and associated cables. Each of the two detector modules is comprised of 12×12 array of 4.14×4.14×20 mm$^3$ LFS crystal pixels with 4.2 mm pitch, one-to-one coupled to SiPM pixels with 75 μm microcells [1]. The overall dimension of the detector module is 50.4×50.4 mm$^2$ eight 18-Ch ASIC chips in the front-end electronic board process the SiPM signals and pass the signal to downstream electronic boards. The ASICs use time-over-threshold (ToT) technique for extracting time and energy information. Similar type TOF-PET modules were recently evaluated [26-28] for diagnostic PET applications. The photo of the demo unit provided by Hamamatsu Photonics is shown in Fig. 2. It should be noted that by default, the demo unit comes with 15 cm cables connecting the two detector boards to the timing PCB, and therefore, it limits the maximum distance between the two detectors to less than 15 cm. It is apparent that TOF PET detectors with expected ~280 ps CTR cannot showcase the impact of TOF information in such a short 15 cm distance. We therefore ordered longer cables to put the detectors at a distance above 25 cm, representing relevant clinical dimensions.

Table 1: Expected vs measured diameter for the spheres in the lesion phantom

| Expected sphere diameter (mm) | Measured sphere diameter (mm) |
|---|---|
| 9.5 | 9.36 |
| 8 | 7.82 |
| 6 | 5.66 |
| 4 | 3.72 |
| 2 | 1.75 |

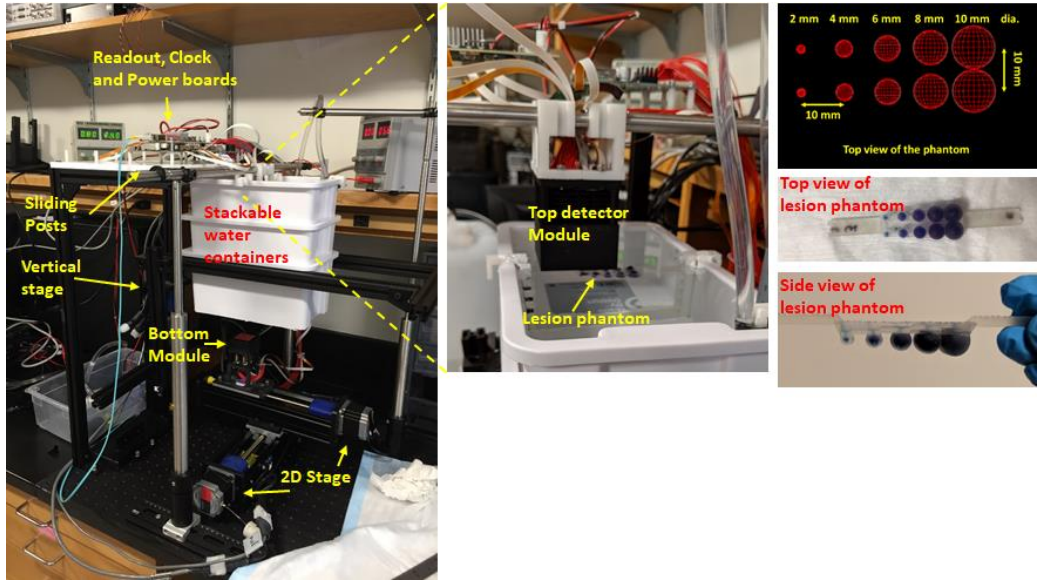

Fig. 3. Photograph of the experimental setup containing detector modules, 2D stages, vertical stage, background water container, and lesion phantom. Also shown is top view of the GATE render of the lesion phantom.

2.2.2   Phantom

In the experimental setup, we used stackable water containers with 24.5×13.5 cm$^2$ base dimension and different water depths up to 25.5 cm. We 3D printed multiple hot sphere phantoms using clear resin with Form3 printer (Formlabs, USA) as shown in Figure 3. To make the hot spheres fillable, we implemented 0.6 mm diameter channels to connect spheres within the phantom to the phantom's top surface. After printing, we measured the volume inside each sphere to verify the phantom with precision of 10 μl for small and up to 1000 μl syringes (Hamilton, USA) for larger spheres. Table 1 shows measured sphere volumes in the phantom. To experimentally investigate the effect of tracer to background ratio on the image quality, we filled three phantoms with 5:1, 10:1, and 20:1 intended activity ratio compared to the background. Note that for better visual, we mixed the radiotracer with food coloring prior to phantom filling. For each experiment, after filling the spheres, we sealed the channels using epoxy sealant. We used $^{18}$F to fill the water container and the hot sphere phantoms. The water container was fixed on a set of construction rails (Thorlabs, USA). This fixture was mounted on a Velmex BSlide (Velmex, USA) vertical motorized stage to control the position of the water tank and the lesion phantom with respect to the coincidence detector modules. We printed a lesion phantom holder to place phantom in different vertical positions inside the water container, and then moved the container vertically to place lesion phantom in its original position with respect to the top detector module. We used a small water pump and a valve to fill and remove active water into and from the water container to have accurate level control for active water and reduce staff exposure and radiation hazard risk. GATE render of the phantom along with photographs of the experimental setup are also shown in Fig. 3.

2.2.3   Experimental procedure

Given the availability of only two TOF-PET detector modules, we mounted one detector module on a horizontal bar atop the rectangular water container and the other module on an X-Y BiSlide motorized stage (Velmex, USA). The top detector can be manually moved in two positions to represent a 5.1×10.2 cm$^2$ detector area. Note that the two positions were selected to cover the hot sphere phantom with enough line of response (LOR) sampling and to avoid long acquisition time imposed by $^{18}$F decay. The bottom detector was moved in a 3×3 position map providing a 15.3×15.3 cm$^2$ detector coverage. Data acquisition was repeated for a total of 2×9 detector positions, where for each top detector position the bottom detector was moved in an step and shoot mode for a fixed acquisition length.

### 2.2.4 Uniformity phantom

In order to calibrate the two detector modules for timing, energy, and uniformity we 3D printed a uniformity phantom with same size of the detector modules placed at the center of hypothetical line connecting the coincidence detectors. Fig. 4 shows uniformity phantom and data acquisition setup. Coincidence data was collected for 7 hours while we maintained the event count rate in both detector module to be less than 3 Mcps by adding more $^{18}$F solution to the phantom every ~45 minutes.

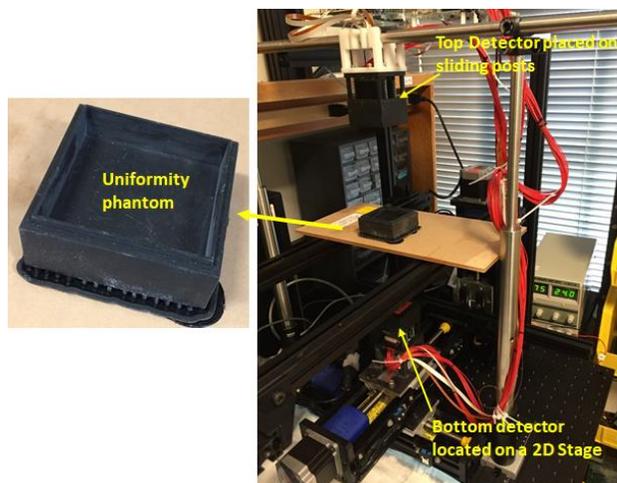

### 2.2.5 Coincidence filter

As part of the demo unit kit, a data collection software was provided to read raw data with Direct Memory Access (DMA). The raw data for each event is packed in 16 bytes and contains absolute event arrival time with 15.0602 ps resolution (47 bits), energy in ToT (8 bits), and board/crystal ID (17 bits). For coincidence filter, the raw data collected in the list-mode should be sorted based on the time stamp. Since the amount of raw data for the uniformity calibration was more than 600 GB, we implemented in software, a 12 stages of fast swapping technique presented in [29] for fast sorting of raw data in coincidence filter. Based on this method, we have also implemented a delayed coincidence filter for correcting random coincidences. A delayed window with 100 ns offset was created for the delay coincidence filter.

Fig. 4. Photograph of the calibration setup used in uniformity, energy, and timing corrections. A 2D scanning stage holds the bottom detector module, while the two-position sliding posts hold the top detector module. At the center plane between the two detectors, the phantom is placed to collect large number of coincidence events for corrections.

### 2.2.6 Energy Correction

At the first stage of data processing the energy correction is required to set the energy window and avoid including scatters and spurious events in the detectors. In practice, we noticed that the detector modules are configured at very low threshold for registering counts to maintain high timing performance and, therefore, the rate of spurious events is higher than in conventional non-TOF detectors. For energy correction we used coincidence data from the acquired uniformity calibration data and obtained energy spectra for each crystal-photodetector pair in the ToT scale. Then we used channel correction factor to overlay photopeak centers in all single crystal-photodetector pair. We used parameters presented in [26] to convert ToT value to keV value, to set the energy window in keV scale instead of ToT scale. Fig. 5 shows energy peak map of detector pixels in two detector modules. After overlaying photo-peaks in the same dataset, energy resolution of detector modules was calculated at 16.7% full-width at half-maximum (FWHM).

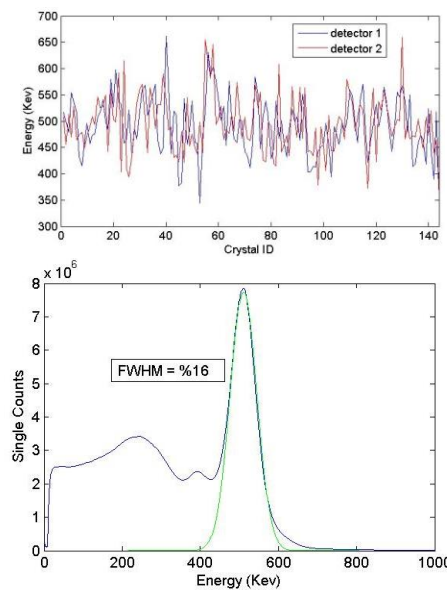

### 2.2.7 Timing Correction

Since there is only minimal amount of scattering media in our uniformity phantom, we used a wide energy window to acquire coincidence data and plot them as 144×144 values in order to find the time-offset for each crystal pair. We used this data to correct timing differences between crystal pairs. Fig. 6a shows time offset between each detector pixel pairs in the two detector modules. The vertical axis is in picoseconds. Note that the cable length between the detector modules to the coincidence unit was different in that, cable length for module 1 was 15 cm, and for module 2 was 100 cm which leads to a ~2.4 ns of arrival time difference. Fig. 6b shows the timing plot for the two detector modules after correcting timing offsets for different pixel pairs with 271 ps FWHM CTR.

Fig. 5. Map of the photopeak position for each of the 144 scintillator pixels of the two detector modules. After energy correction for the two detectors energy resolution of 16.7% was achieved.

### 2.2.8 Uniformity Correction

Uniformity correction was performed based on the same data collected for timing correction after we stored each detector pair counts in a 144×144 matrix as the reference for uniformity correction with average of 645 counts per each pair (total of ~13.4 M coincidence counts).

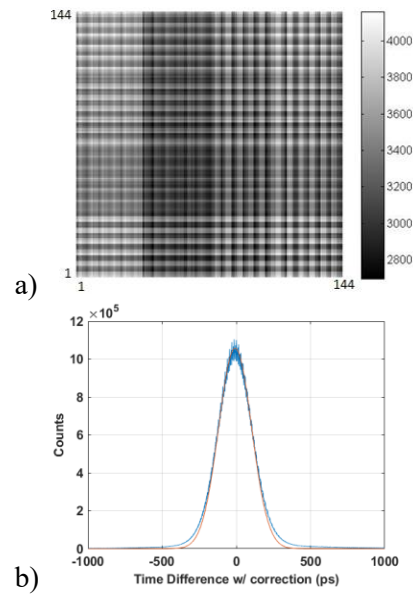

Fig. 6. Time offset between each crystal pair of the two detector modules, a) before and b) after correction with 271 ps FWHM CTR.

## 2.3 Image reconstruction and quality evaluation

### 2.3.1 Image reconstruction

While acknowledging superior performance of iterative and statistically-based image reconstruction methods, for simplicity and consistency we have implemented a simple 3D back-projection method with TOF information [15]. In this method, we back-projected a Gaussian curve with 271 ps FWHM into the image matrix. The image matrix comprised of 256×256×256 voxels with each voxel representing a 1 mm$^3$ volume. However, the small image voxel when used with 4.2 mm detector pixels, leads to speckle artefact in the reconstructed image. To minimize this artefact we employed oversampling of the interaction position in X-Y directions across the detector pixel surface [15]. By using oversampling in image reconstruction, the computation time increases linearly by the oversampling factor in X and Y directions, which can be minimized by employing GPU instead of CPU. In all reconstructed images in this work, we used 10x oversampling factor in each dimension of the crystal pixel surface creating 100 LORs from one LOR in the list-mode data. Image reconstruction was performed using single CPU core.

### 2.3.2 Image quality metrics

After reconstruction, we evaluated the image quality using Contrast Recovery Coefficient (CRC), Signal to Noise Ratio (SNR), and Contrast to Noise Ratio (CNR). CRC is described in NEMA standard to measure the quality of hot rods in the scan of micro-Derenzo phantom. It represents how much of the real activity is actually recovered in the reconstructed image. We modified the CRC to reflect the use of hot spheres instead of rods (with no slice summation) as follow:

$$CRC = \frac{\frac{Mean_{HotSphere}}{Mean_{Background}} - 1}{\frac{Activity_{HotSphere}}{Activity_{Background}} - 1}$$

where "*Activity*" refers to the activity of the phantom and background in the same region of interest (ROI) where the "Mean" is calculated. CNR shows the detectability of the hot spheres and is defined as the ratio of the image contrast to the noise in the background:

$$CNR = \frac{Mean_{HotSphere} - Mean_{Background}}{Std_{Background}}$$

and SNR is calculated using:

$$SNR = \frac{\frac{Mean_{HotSphere}}{Mean_{Background}} - 1}{\sqrt{\left(\frac{Std_{HotSphere}}{Mean_{HotSphere}}\right)^2 + \left(\frac{Std_{Background}}{Mean_{Background}}\right)^2}}$$

where "*Std*" refers to standard deviation.

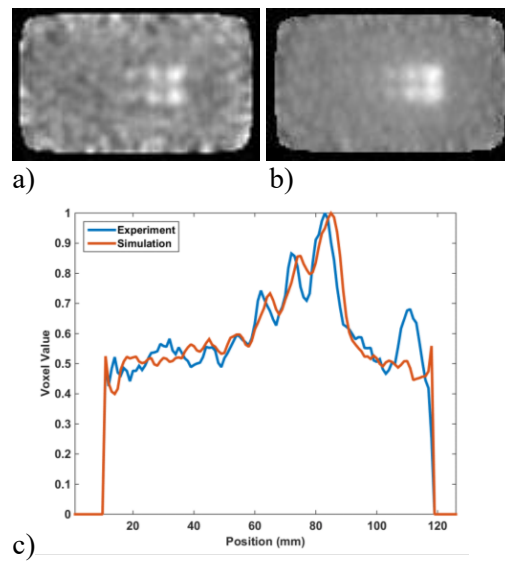

Fig. 7. Experimental (a) and simulation (b) results of lesion phantom with ~10:1 lesion to background activity in a 25 cm deep background water. c) shows vertical line profile through upper line of the hot spheres in the lesion phantom. Acquisition time was 180 seconds.

# 3 Results

The image matrix size in image reconstruction is 256×256 with 1 mm size in all directions. All reconstructed images are shown in coronal view of the plane containing the hot lesion spheres, to be most consistent with the surgeon view in an intraoperative procedure.

## 3.1 Simulation vs experiment

For the main experiment we scanned lesion phantom with 9.1:1 lesion to background ratio in 180 seconds for each of 18 detector position steps. The overall scan took 61 minutes including the detector module re-positioning for all the 18 positions. After decay correction, which was implemented by having the 18$^{th}$ position as reference and cutting coincidence counts from the acquired data for position steps 1 to 17, so that the reconstructed scan time ranged from 125 s for the first detector position to 180 s for the 18$^{th}$ position. We then compared the result with 180 s simulation with the same amount of activity at the end of the experiment. Results are shown in Fig. 7 and Table 2 after applying energy, TOF, uniformity, and sensitivity correction for the experimental data and sensitivity correction for simulated data.

## 3.2 Effect of detector timing resolution

To illustrate the effect of timing resolution of the detector modules we compared the simulation result of a detector system with 271 ps to that of a 100 ps FWHM CTR. Fig. 8 shows image quality improvement due to the better CTR as manifested by clearly resolved 6 mm Dia. spheres well above the background noise in the line profile. Image quality metrics are presented in Table 3.

## 3.3 Effect of lesion to background ratio

To investigate the effect of uptake ratio on the image quality, we carried out experiments with a fixed 60 second acquisition time and with 4.5:1, 9.1:1 and 18.9:1 lesion/background phantoms. All measurements were performed on the same day and, and thus the activity at the end of each experiment was different. Therefore, we trimmed all experimental data according to the activity of background water at the end to the shortest scan which was the one with 4.5 uptake ratio. The results are presented in Fig. 9 and Table 4 where as expected the phantom with higher uptake ratio results in a better image quality.

## 3.4 Effect of background water depth

In addition, in order to study the effect of scattering media on the image quality, we performed measurements using the 9.1:1 phantom at different background water depth. Note that we only moved the bottom detector module and kept the lesion phantom in the same location with respect to top detector module. Similar to the previous experiment in the section, we trimmed experimental data according to the activity of background water at the end of each step to the shortest collected data point. Results are shown in Fig. 10 and Table 5.

Table 2: Image quality parameters comparing simulation and experimental study

| | | Sphere Diameter (mm) | | | | |
|---|---|---|---|---|---|---|
| | | 10 | 8 | 6 | 4 | 2 |
| CNR | Experiment | 7.489 | 6.227 | 3.485 | 1.807 | 0.443 |
| | Simulation | 17.116 | 12.968 | 7.342 | 3.368 | 0.986 |
| CRC | Experiment | 0.065 | 0.054 | 0.03 | 0.016 | 0.004 |
| | Simulation | 0.075 | 0.057 | 0.032 | 0.015 | 0.004 |
| SNR | Experiment | 0.037 | 0.043 | 0.027 | 0.018 | 0.003 |
| | Simulation | 0.444 | 0.379 | 0.31 | 0.188 | 0.033 |

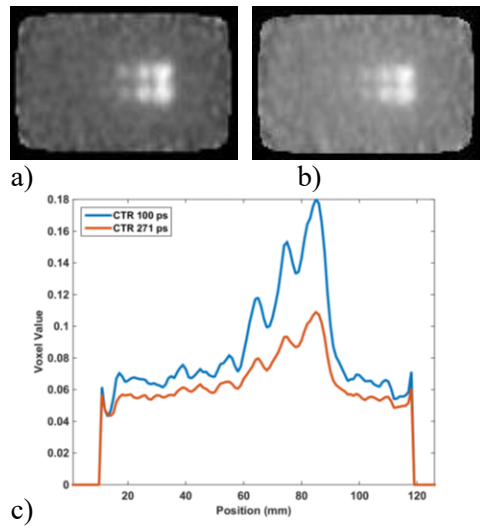

Fig. 8. Simulation results showing the impact of a) 100 ps CTR on image quality compared to b) 271 ps FWHM CTR detector. Corresponding line profiles are shown in c).

Table 3: Image quality parameters comparing timing resolution effect

| | CTR Value | Sphere Diameter (mm) | | | | |
|---|---|---|---|---|---|---|
| | | 10 | 8 | 6 | 4 | 2 |
| CNR | 100 ps | 19.475 | 14.781 | 8.046 | 2.73 | 0.012 |
| | 200 ps | 17.116 | 12.968 | 7.342 | 3.368 | 0.986 |
| CRC | 100 ps | 0.14 | 0.106 | 0.058 | 0.02 | 0 |
| | 200 ps | 0.075 | 0.057 | 0.032 | 0.015 | 0.004 |
| SNR | 100 ps | 0.665 | 0.524 | 0.371 | 0.166 | 0.001 |
| | 200 ps | 0.444 | 0.379 | 0.31 | 0.188 | 0.033 |

Table 4: Image quality parameters comparing lesion to background ratio (LBR)

| | LBR | Sphere Diameter (mm) | | | | |
|---|---|---|---|---|---|---|
| | | 10 | 8 | 6 | 4 | 2 |
| CNR | 5 | 1.511 | 2.107 | 0.11 | 0.402 | 0.301 |
| | 10 | 6.921 | 5.121 | 2.48 | 1.17 | 1.577 |
| | 20 | 7.264 | 5.606 | 2.183 | 0.528 | 0 |
| CRC | 5 | 0.018 | 0.026 | 0.001 | 0.005 | 0.004 |
| | 10 | 0.069 | 0.051 | 0.025 | 0.012 | 0.016 |
| | 20 | 0.125 | 0.096 | 0.038 | 0.009 | 0 |
| SNR | 5 | 0.005 | 0.007 | 0 | 0.002 | 0.001 |
| | 10 | 0.018 | 0.02 | 0.01 | 0.006 | 0.007 |
| | 20 | 0.028 | 0.023 | 0.01 | 0.003 | 0 |

Table 5. Image quality parameters comparing different background water depths

| | Depth (cm) | Sphere Diameter (mm) | | | | |
|---|---|---|---|---|---|---|
| | | 10 | 8 | 6 | 4 | 2 |
| CNR | 4 | 9.765 | 7.058 | 2.653 | 1.059 | 0.214 |
| | 8 | 7.282 | 4.796 | 1.872 | 0.485 | 0 |
| | 12 | 4.855 | 3.576 | 1.941 | 1.268 | 0.033 |
| | 25 | 5.614 | 3.919 | 1.926 | 0.67 | 1.1 |
| CRC | 4 | 0.119 | 0.086 | 0.032 | 0.013 | 0.003 |
| | 8 | 0.083 | 0.055 | 0.021 | 0.006 | 0 |
| | 12 | 0.077 | 0.056 | 0.031 | 0.02 | 0.001 |
| | 25 | 0.072 | 0.05 | 0.025 | 0.009 | 0.014 |
| SNR | 4 | 0.073 | 0.063 | 0.026 | 0.011 | 0.002 |
| | 8 | 0.066 | 0.052 | 0.021 | 0.006 | 0 |
| | 12 | 0.034 | 0.028 | 0.017 | 0.011 | 0 |
| | 25 | 0.013 | 0.012 | 0.005 | 0.003 | 0.003 |

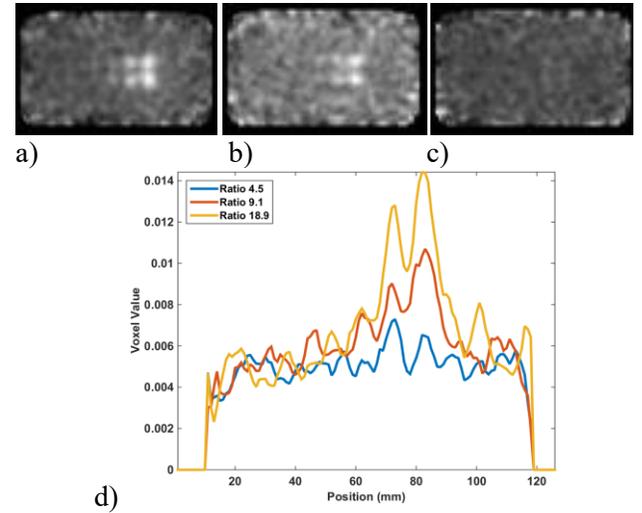

Fig. 9. Experimental results showing the impact of lesion to background ratio on image quality when using flat panel PET detectors with 271 ps FWHM CTR. lesion to background ratio of 18.9, 9.1, and 4.5 are shown in a), b), and c), respectively. Superimposed line profiles are shown in d).

## 4  Discussion & Conclusion

We have demonstrated that the ready to use compact TOF-PET technology from Hamamatsu Photonics can be adapted to our intraoperative PET imager concept. Using two-module demonstration kit we have developed a correction procedure that allows calibration of the detector modules and correction for non-uniformities in response. We studied the effect of tracer uptake, background depth, detector CTR after validating the experimental data against simulation setup.

For simplicity we implemented simple 3D oversampled back projection with TOF information for image reconstruction. With short acquisition time of ~60 seconds and no iterative reconstruction algorithm, the image quality is limited, and 6 mm spheres are only detectable when the 18:1 activity ratio was used. It is known that limited angle PET can benefit from iterative reconstruction algorithms, however these techniques are typically computationally intensive and require GPU-based implementation otherwise will lead to delayed reconstructed image which may not be useful for an intraoperative procedure. The data presented here shows that some of the artefact due to limited angle data can be reduced when using detector with improved CTR. This is an active area of research by our group among others, and we plan to implement improved detector technology with both improved DOI [30, 31] and CTR [32-34] in future work. Also noteworthy is that by using 4.2 mm detector pixel pitch and no depth of interaction information, detection of <4 mm hot spheres with short acquisition is not practical. Despite the limited detector technology and limited detector coverage, and suboptimal image reconstruction technique, our results are very encouraging in that by implementing improved TOF-PET technologies, one can aim at improved image quality which may lead to provide an alternative solution to intraoperative imaging and ultimately improve the patient outcome.

**Acknowledgements**
The authors would like to thank Martin Janecek and Ardavan Ghasemi at Hamamatsu for lending us the TOF-PET kit for evaluation. Authors

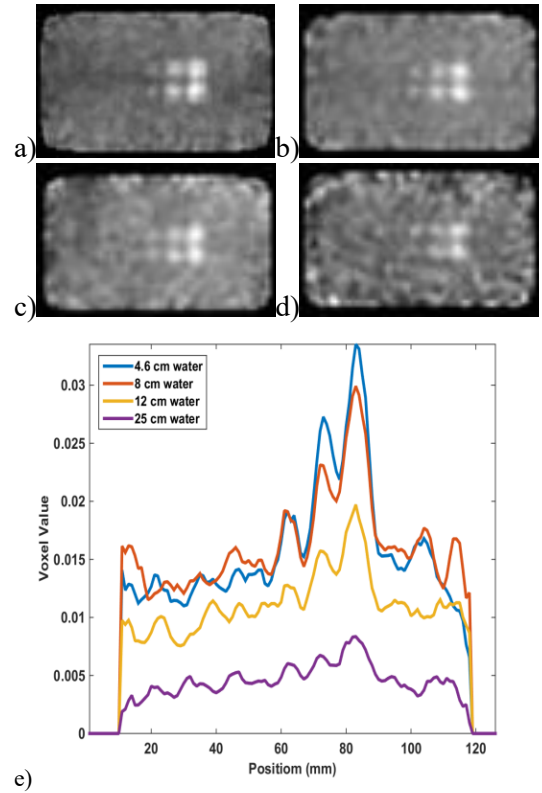

Fig. 10. Experimental results showing the impact of background water depth on the image quality when using flat panel PET detectors with 271 ps FWHM CTR. Background water depth of 4.6 cm, 8 cm, 12 cm, and 25 cm are shown in a, b, c, and d, respectively. e) shows the superimposed line profile through one pixel row in the reconstructed images.

thank Beca Vittum, Hushan Yuan, and Abhishikth Devabhaktuni for their help with the experimental preparation.